\documentclass{aa}  

\usepackage{graphicx}
\usepackage{txfonts}
\usepackage{soul}

\usepackage{xcolor}

\begin{document} 

   \title{Effects of coronal mass ejection orientation on its propagation in the heliosphere}


   \author{K. Martini\'{c}
          \inst{1}
          \and
          M. Dumbovi\'{c}\inst{1}
          \and
          J. \v{C}alogovi\'{c}\inst{1}
          \and
          B. Vr\v{s}nak\inst{1}
          \and
          N. Al-Haddad\inst{3}
          \and
          M. Temmer\inst{2}
          }

   \institute{Hvar Observatory, Faculty of Geodesy, University of Zagreb, Zagreb, Croatia\\
              \email{kmartinic@goef.hr}
         \and
              Institute of Physics, University of Graz, Graz, Austria
        \and
              University of New Hampshire, Space Science Center, Durham, USA
             }

   \date{Received September 15, 1996; accepted March 16, 1997}

 
  \abstract
   {In the scope of space weather forecasting, it is crucial to be able to more reliably predict the arrival time, speed, and magnetic field configuration of coronal mass ejections (CMEs). From the time a CME is launched, the dominant factor influencing all of the above is the interaction of the interplanetary CME (ICME) with the ambient plasma and interplanetary magnetic field.}
   {Due to a generally anisotropic heliosphere, differently oriented ICMEs may interact differently with the ambient plasma and interplanetary magnetic field, even when the initial eruption conditions are similar. For this, we examined the possible link between the orientation of an ICME and its propagation in the heliosphere (up to 1 AU).}
   {We investigated 31 CME-ICME associations in the period from 1997 to 2018. The CME orientation in the near-Sun environment was determined using an ellipse-fitting technique applied to single-spacecraft data from SOHO/LASCO C2 and C3 coronagraphs. In the near-Earth environment, we obtained the orientation of the corresponding ICME using in situ plasma and magnetic field data. The shock orientation and nonradial flows in the sheath region for differently oriented ICMEs were investigated. In addition, we calculated the ICME transit time to Earth and drag parameter to probe the overall drag force for differently oriented ICMEs. The drag parameter was calculated using the reverse modeling procedure with the drag-based model.}
   {We found a significant difference in nonradial flows for differently oriented ICMEs,
    whereas a significant difference in drag for differently oriented ICMEs was not found. }
   {}


   \maketitle
%

\section{Introduction}

A coronal mass ejection (CME) is a large-scale ejection of plasma and magnetic field from the solar corona into the interplanetary medium. When it reaches Earth, it can cause large disturbances in the near-Earth environment (i.e., it can trigger geomagnetic storms). It is relatively widely accepted that CMEs consist of a so-called flux rope (FR) structure (\citealp{Chen1996}; \citealp{Bothmer1998}; \citealp{Moore2001}) that may drive sheaths and shocks. An FR, in its simplest form, is a cylindrical structure in which a poloidal magnetic field component rotates about an axial magnetic field component that follows the central axis of the cylinder \citep{LUNDQUIST1950}.

Coronal mass ejections have been observed remotely with white-light coronagraphs. A CME FR reconstruction can be performed using stereoscopic coronagraph images. \cite{Thernisien2006} developed a 3D model for CME FR reconstruction, referred to as the graduated cylindrical shell (GCS) model, in which an FR is represented as a "hollow croissant" consisting of two conical legs and a curved front. One of the six main parameters to fully describe the FR in the GCS reconstruction is tilt. The tilt of an FR is defined as the angle between the solar equator and the central axis of the FR. It is measured from solar west to solar north (positive values) and from solar west to solar south (negative values). Defined in this way, the tilt essentially gives the inclination of the CME with respect to the solar equator. Another way to determine the inclination of a CME is based on a 2D CME reconstruction, first proposed by \cite{Chen1997}, where the observed CME front is represented with an ellipse. In this model, changing the position of the ellipse, the length of the axes, and the inclination of the major axis of the ellipse can account for the angular width and inclination of the CME (\citealp{Krall2006}; \citealp{Byrne2009}; \citealp{Martinic2022}). \cite{Martinic2022} showed that GCS and ellipse fitting give comparable results for the inclination of CMEs when using remote data from coronagraphs aboard the SOHO and STEREO spacecraft for 22 Earth-directed events. 
 
Commonly, there is a distinction between the CMEs observed remotely in the corona and the interplanetary CMEs, or ICMEs, measured in situ by spacecraft. Recently, however, in situ measurements of CMEs in the upper corona and innermost heliosphere taken with the Parker Solar Probe and Solar Orbiter have caused this traditional distinction between CMEs and ICMEs to become less clear. In this study, we use the term "ICME" in the context of in situ measurements and interplanetary interaction with the ambient; for the rest, the "CME" term is used.

Typically, the three-part structure (the shock, the sheath, and the magnetic obstacle) can be well-measured as the spacecraft passes an ICME. First, a fast-forward shock front is usually detected, characterized by an abrupt increase in magnetic field, solar wind speed, and temperature. After the shock front, a so-called ICME sheath region is measured. This is a special case of plasma sheaths where both expansion and propagation properties are observed \citep{Siscoe2008}. The ICME sheaths are turbulent and compressed, as evidenced by elevated values and strong fluctuations of the magnetic field, density, velocity, and plasma beta parameter \citep{Kilpua2017}. After the sheath is the driver, the FR part of the ICME, that is, the magnetic obstacle (MO). A subset of well-defined MOs is called a magnetic cloud (MC), which is characterized by a smoothly rotating magnetic field, decreased plasma beta parameter, and decreased temperature \citep{Burlaga1991}. As a first approximation, and based on their chirality and orientation, ICMEs can be classified into eight basic types, as described in \cite{Bothmer1998}, \cite{Mulligan1998}, and recently by \cite{Palmerio2018}. Four of these eight types are low-inclined ICMEs, and the remaining four are high-inclined ICMEs. 

Three forces are active during different CME propagation phases. In the early acceleration phase, the Lorentz and gravitational forces compete with each other. Later, the magnetohydrodynamic (MHD) drag force from the solar wind acts on the CME. Observations have shown that CMEs faster than the solar wind slow down, while CMEs slower than the solar wind accelerate (\citealp{Sheeley1999}; \citealp{Gopalswamy2000}; \citealp{vrsnak2004}; \citealp{Manoharan2006}).

\begin{figure}
   \centering
   \includegraphics[width=\hsize]{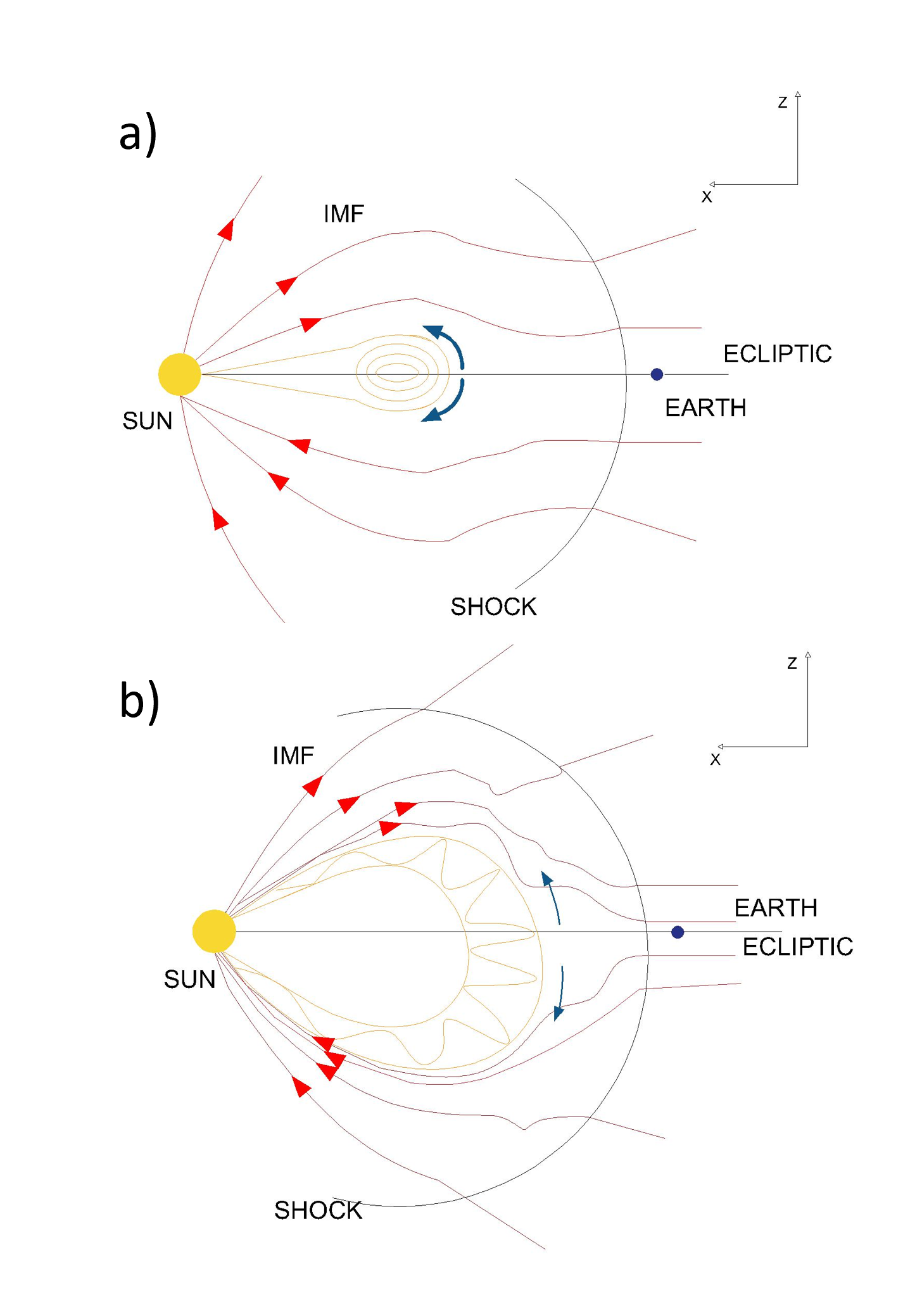}
      \caption{Idealized IMF in the meridional plane, xz-plane of GSE coordinate system, and its interaction with embedded ICME with low inclination (upper panel) and high inclination (bottom panel). The NRF is shown with blue arrows where its width and length suggest the pronouncement of the plasma flows in front of the embedded ICME. The figure is adapted from \cite{Martinic2022}.
              }
         \label{Fig1}
   \end{figure}

Drag in interplanetary space (MHD drag) is not primarily caused by viscosity and particle collisions but is rather related to the interaction of the ICME with the surrounding magnetic field, such as MHD waves \citep{Cargill1996} and magnetic field draping \citep{Gosling1987}, as described in \cite{Martinic2022}. Interplanetary CMEs interact with the surrounding plasma and magnetic field as they propagate in the heliosphere. For fast ICMEs embedded in the slow ambient plasma, accelerations and deflections of the ambient plasma occur in front of the ICME FR part. Due to the high electrical conductivity, the ambient solar wind cannot easily penetrate the magnetized ICME structure, but it is accelerated and deflected around the obstacle. This occurs in an ICME sheath region and is particularly pronounced near the ICME FR part. A direct consequence of this plasma motion is the draping of the IMF around the ICME FR. Apart from the relative velocity between the ICME and the surrounding solar wind, the draping pattern depends strongly on the size and shape of the ICME and on the configuration of the surrounding magnetic field (\citealp{Gosling1987}; \citealp{McComas1988}; \citealp{McComas1989}). Consequently, for differently oriented ICMEs, even if embedded in similar configurations of the ambient magnetic field and solar wind, one might expect a different plasma flow and consequently a different draping pattern, as theorized by \cite{Martinic2022}. Figure \ref{Fig1} shows a low-inclination ICME in panel (a) and a high-inclination ICME embedded in the surrounding magnetic field in panel (b). Only the meridional plane, the xz-plane of the Geocentric Solar Ecliptic (GSE) coordinate system, is shown in Figure \ref{Fig1}, and one should consider the Parker spiral (i.e., the Parker spiral configuration of the magnetic field in the xy-plane). In the case of ICMEs with high inclination, more draping occurs due to the interaction with the broader extent of the ICME front. The blue arrows in Figure \ref{Fig1} schematically represent the plasma flows in front of the obstacle. Due to the larger pressure gradient associated with the pileup of the magnetized solar wind, the ambient plasma is expected to pass the obstacle more easily in the direction in which the extent of the obstacle is smaller. Thus, in an ICME with low inclination, the plasma flow in the xz-plane of the GSE coordinate system is more pronounced than in an ICME with high inclination. In contrast, for an ICME with high inclination, one would expect more pronounced plasma flows in the yz-plane (into and out of the plane shown in Figure \ref{Fig1}). The ambient field that is draped eventually slides past the obstacle. This process should be more efficient for an ICME with a low inclination since the expansion in the xz-plane is smaller, and the ICME can push the draped field around the obstacle more easily than an ICME with high inclination.

%

\cite{Vandas1995} and \cite{Vandas1996} studied the propagation of two MCs, one low inclined and one high inclined, represented by Lundquist's cylindrical force-free solution \citep{LUNDQUIST1950} in the inner heliosphere using the 2.5D MHD model. Details of this model can be found in \cite{Wu1979} (2D) and \cite{Wu1983} (2.5D). They found that the propagation of these MCs does not depend on the inclination of their axes with respect to the ecliptic plane (one lies in the ecliptic, and the other has an axis perpendicular to it). The MHD model used in these studies was confined to the solar equatorial plane and therefore does not provide a complete 3D MHD representation. In order to provide a better forecast of ICME arrivals, the influence of field line draping and associated nonradial flows (NRFs) on the ICME propagation from the observational perspective needs to be investigated on a statistically relevant sample of events. To our knowledge, this influence was first studied by observation in \cite{Martinic2022}. In this present study, we extend the data sample to provide better statistical coverage and investigate the effects of NRFs and field line draping on the propagation behavior of the CME. In Section \ref{Data and method}, we describe the method by expanding on the study by \citet{Martinic2022}. We highlight several dynamical features used to study the interaction between differently oriented ICMEs and the environment. In terms of the plasma flows in front of the ICME FR, we studied NRFs and shock orientation; and in terms of the overall drag, we studied drag parameter and ICME transit time. The main findings are presented in Section \ref{Results and discussion}, and our conclusions are in Section \ref{Summary and conclusions}.

\section{Data and method}
\label{Data and method}

   \begin{figure*}
   \centering
   \includegraphics[width=2\columnwidth]{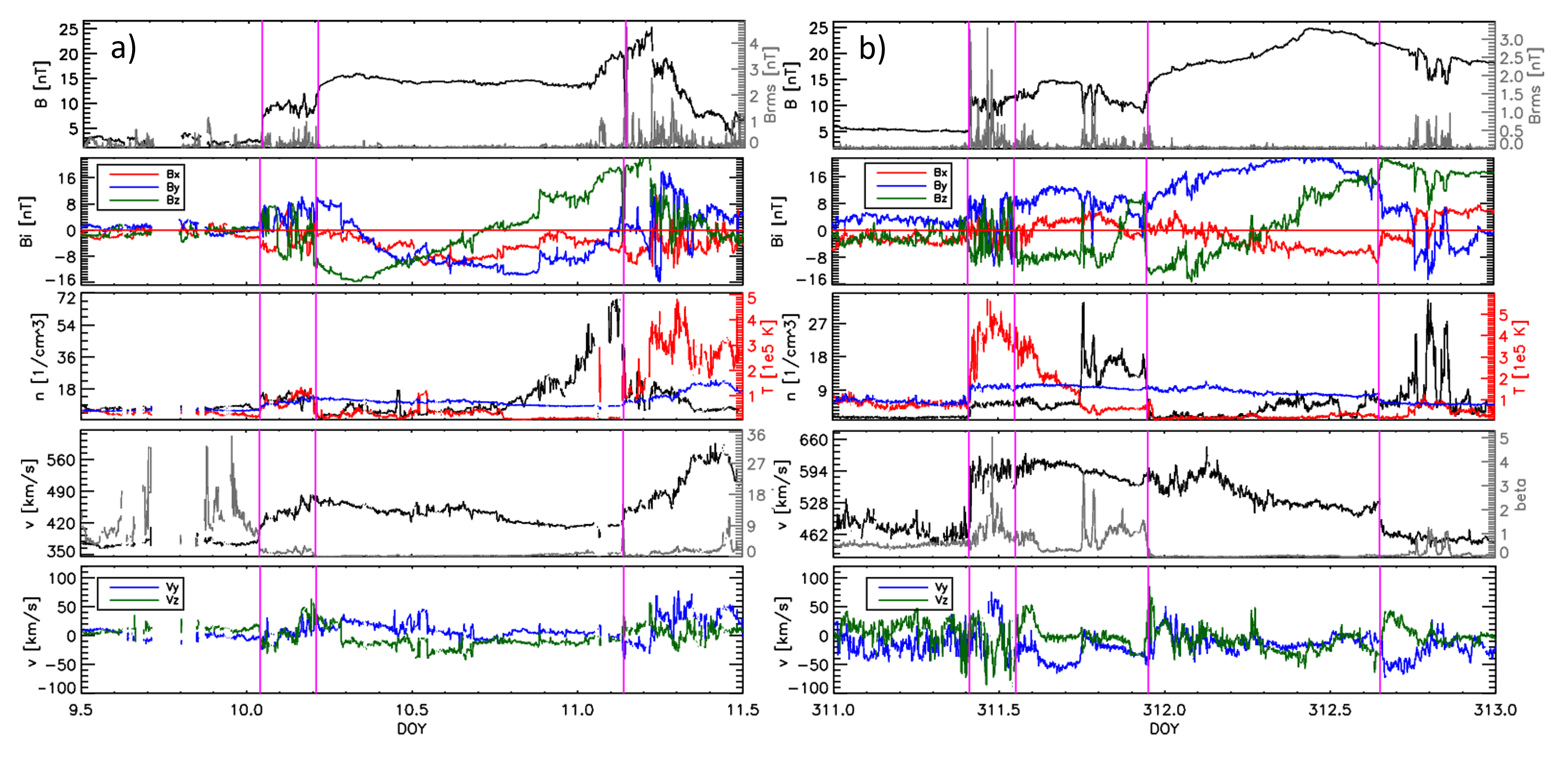}
   \caption{Interplanetary CME measured in situ on 10 January 1997  (left panels) and 3 November 2000 (right panels). From top to bottom, the following parameters are shown: Magnetic field magnitude in black and magnetic field fluctuations in gray (right scale); GSE magnetic field components (red, $B_x$; blue, $B_y$; green, $B_z$); proton density in black, temperature in red, and expected temperature in blue; solar wind speed in black and plasma beta parameter in gray; GSE velocity components (blue, $B_y$; green, $B_z$). From left to right, the vertical magenta lines mark the shock arrival, the end of the clear sheath, and the MO end time. In the right panels, the end of the clear sheath part does not coincide with the MO onset time, and there is an additional vertical magenta line present.}
              \label{Fig2}
    \end{figure*}

   \begin{figure}
   \centering
   \includegraphics[width=\hsize]{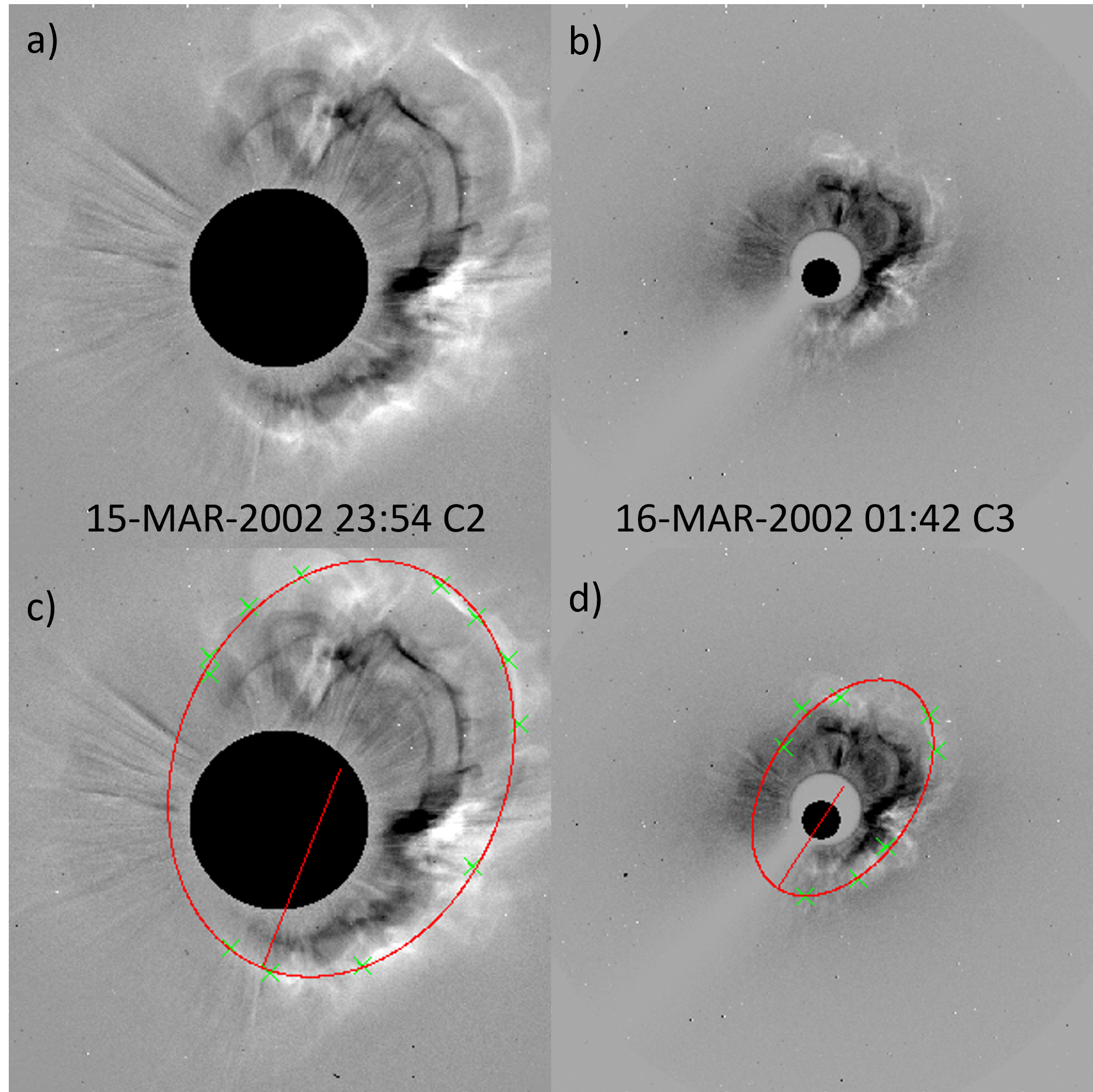}
      \caption{Coronal mass ejection that occurred on 15 March 2002. The upper panels show the running difference images in LASCO-C2 (left) and LASCO-C3 (right). The bottom panels show the corresponding ellipse fitting. The ellipse is indicated with a red line, whereas green crosses mark the points outlined on the CME front used to obtain the fit.
              }
         \label{Fig3}
   \end{figure}
%

We searched for associated CME-ICME pairs from 1996 to 2020. The lists we used to create our sample can be found in the following studies: \citealp{Nitta2017} (abbr. NM), \citealp{Palmerio2018} (abbr. P), \citealp{Temmer2021} (abbr. T), and \citealp{Xie2021} (abbr. X).

In total, 113 CME-ICME pairs were found, but only 31 were used in our analysis. Most events were excluded for two reasons: insufficiently developed sheath region (32 excluded) and unclear MO boundary determination (30 excluded).  The former relates to missing signatures of a clear sheath region ahead of the MO \citep[for a discussion of CMEs with and without sheath regions, see][]{Salman2020}. As highlighted in \cite{Kilpua2017}, the sheath thickness depends on the velocity and physical properties of the driving MO and the ambient solar wind, but sheath thickness has also been shown to increase from the nose toward the flanks. Unclear MO boundary determination is related to the subjectivity in determining the boundaries of the MO. There are some MO examples where there are clearly multiple rotations of the same or different magnetic field components, and in such cases, it is not straightforward to establish the MO boundaries and associate the example with a simple FR categorization of eight types. Other reasons why some of the events were excluded are as follows: faint CME front and multiple eruptions within the LASCO field of view (11 excluded); possible ICME interactions with other ICMEs or high-speed streams (4 excluded); no clear magnetic field rotation, that is ejecta-ICME, (1 excluded); no in situ data (1 excluded); possible incorrect CME-ICME association (1 excluded); and inconsistent dominant inclination derived from remote observations and in situ measurements (2 excluded). Ultimately, 31 CME-ICME pairs in the period from 1997 to 2018 with clear MO signatures were left.

\subsection{Dominant inclination determination}
 We derived the dominant inclination for the CME-ICME pairs from both the remote and in situ data. For the remote data, we used SOHO/LASCO \citep{Brueckner1995} coronagraph images and performed an ellipse fit. This method assumes that the outer edge of the (partial) halo CME can be represented by an ellipse whose major axis inclination indicates the dominant inclination of the CME. An example of the application of the ellipse-fitting technique to event number eight is shown in Figure \ref{Fig3}. The top row shows running difference images in the LASCO-C2 and LASCO-C3 field of view (FOV). In the bottom row, the ellipse fitting is overlaid with a red line.

In situ data was obtained from the WIND and ACE space probes, available through the OMNI database \citep{King2005}. The dominant inclination from the in situ data was derived from the rotation of the magnetic field components in the MO part of the ICME using the GSE system. If the rotation of the $B_z$ component was observed to change sign but the $B_y$ component retained its sign, we considered the event to be a dominantly low-inclined event (see Figure \ref{Fig2}). On the other hand, if a sign change was observed in the $B_y$ component but the $B_z$ component remained the same throughout the MO, the event was considered to be dominantly high inclined. We divided all events into eight basic categories. Four of these eight categories are dominantly high inclined (ESW, ENW, WSE, and WNE), and the other four are dominantly low inclined (SWN, NWS, SEN, and SWN). Here, E stands for east, W for west, N for north, and S for south. The ESW type has an axis directed toward the south and a helical field rotating from east to west. The ENW type has the same helical field rotation, but the axial field is directed toward the north. The same applies to the others. The results of the classification are shown in Table \ref{table:2}. \cite{Al-Haddad2013} found that FR reconstruction shows different inclinations for different FR reconstruction techniques, and this varies greatly with the MO boundary set. This is the reason why we only distinguish between dominantly high- and dominantly low-inclined events, rather than deriving the exact inclination for each event \citep[see ][]{Martinic2022}.

In summary, we divided all events into two groups: events with predominantly low inclination and those with predominantly high inclination. Events with predominantly low inclination are those with an inclination of less than 40$^\circ$, as determined from the ellipse fit, and with a rotation in the $B_z$ magnetic field component (ESW, ENW, WSE, and WNE), as observed in situ. Events with predominantly high inclination are those with an inclination greater than 45$^\circ$, as determined from the ellipse fit, and with rotation in the $B_y$ magnetic field component (SWN, NWS, SEN, and NES), as seen in situ. We considered the events with an inclination between 40$^\circ$ and 45$^\circ$ to be intermediate inclination events and did not include them in the analysis. 

For two CME-ICME pairs that were excluded, we found inconsistencies in the dominant inclination inferred from the in situ and remote data. \cite{Xie2021} showed that 25$\%$ of the events studied had a rotation of more than 40$^\circ$ from the near-Sun to L1. They also showed that 56$\%$ of these events exhibited rotation in the STEREO/SECCHI-COR2 FOV (i.e., in the mid-corona). \cite{Isavnin2013} showed that about one-third of the events studied showed a change in inclination from predominantly low to high, or vice versa. In our sample of 33 events, we found only two events where this was true. This could be due to the fact that we excluded over 30 CME-ICME pairs because of ambiguous rotation of the magnetic field components within the MO part of the ICME. Of the remaining 31 events, 19 are dominantly low inclined, while 12 are dominantly high inclined. These 31 CMEs are listed in Table \ref{table:1}, and their interplanetary counterparts, ICMEs, are listed in Table \ref{table:2}. The first column of Table \ref{table:1} shows the event number accompanied by an abbreviation indicating which study the CME-ICME association was taken. The second column shows the first C2 appearance time as reported in the SOHO/LASCO CME catalog.\footnote{\url{https://cdaw.gsfc.nasa.gov/CME_list/}} The third and fourth columns show the time at which the ellipse fit reconstruction was performed in the LASCO-C2 and LASCO-C3 FOV, respectively. This is followed by the columns showing the obtained tilt, in LASCO-C2 FOV and LASCO-C3 FOV, respectively. The last column shows whether the event is dominantly high or dominantly low inclined, as obtained from the ellipse fit in the LASCO-C2 and LASCO-C3 FOV. The letter "L"\ indicates that the event is dominantly low inclined and that the average of the absolute tilt values obtained from the ellipse fit reconstruction in LASCO-C2 and LASCO-C3 FOV is less than 40$^\circ$. The letter "H" indicates that the event is dominantly high inclined. Analogously, such events are those whose average absolute tilt values are higher than 45$^\circ$.

%
\begin{table*}           
\label{table:1}      
\centering          
\begin{tabular}{ l l l l r r r}     
 Nr. & First C2 Appearance & Ellipse Fit in C2 & Ellipse Fit in C3 & Tilt C2 $[^\circ]$ & Tilt C3 $[^\circ]$& Inclination \\
\hline\hline
1$^{X}$&1997-01-06 15:10& no data &1997-01-07 01:59& &3&L\\
2$^{X}$&1997-10-06 15:28&1997-10-06 18:44&1997-10-07 02:31&20&20&L\\
3$^{X}$&1997-11-04 06:10&1997-11-04 06:42&1997-11-04 09:40&-87&74&H\\
4$^{X}$&1998-01-02 23:28&1998-01-03 01:28&1998-01-03 03:40&-51&-89&H\\
5$^{X}$&2000-08-09 16:30&2000-08-09 17:04&2000-08-09 21:18&33&30&L\\
6$^{X}$&2000-11-03 18:26&2000-11-03 22:26&2000-11-04 01:42&-33&-27&L\\
7$^{X}$&2001-04-26 12:30&2001-04-26 13:29&2001-04-26 14:15&35&8&L\\
8$^{X}$&2002-03-15 23:00&2002-03-15 23:52&2002-03-16 01:42&71&67&H\\
9$^{X}$&2002-04-15 03:50&2002-04-15 04:50&2002-04-15 05:18&21&24&L\\
10$^{X}$&2003-08-14 20:06&2003-08-14 21:53&2003-08-15 02:40&55&50&H\\
11$^{X}$&2005-05-13 17:12&2005-05-13 17:22&no data&54& &H\\
12$^{T,X}$&2008-12-12 08:54&2008-12-12 11:54&2008-12-12 15:42&-35&-37&L\\
13$^{T,X}$&2010-04-03 10:33&2010-04-03 10:50&2010-04-03 12:42&6&-18&L\\
14$^{T,NM,X}$&2010-06-16 14:54&2010-06-16 20:06&no data&-30& &L\\
15$^{P,T,X}$&2011-06-02 08:12&2011-06-02 08:48&2011-06-02 09:54&-57&-57&H\\
16$^{P,X}$&2011-09-14 00:00&2011-09-14 01:36&2011-09-14 03:06&-15&-5&L\\
17$^{P,T,X}$&2011-10-22 01:25&2011-10-22 02:24&2011-10-22 04:18&-53&-55&H\\
18$^{P,T}$&2012-01-19 14:36&2012-01-19 15:48&2012-01-19 16:42&-2&-20&L\\
19$^{P,T,X}$&2012-06-14 14:12&2012-06-14 14:36&2012-06-14 16:18 &18&11&L\\
20$^{P,T,X}$&2012-07-12 16:48&2012-07-12 17:24&no data&50& &H\\
21$^{P,T,NM,X}$&2012-10-05 02:48&2012-10-05 05:24&2012-10-05 08:42&45&45&H\\
22$^{T,X}$&2012-11-09 15:12&2012-11-09 16:00&2012-11-09 18:20&31&22&L \\
23$^{P,X}$&2013-01-13 12:00&2013-01-13 15:54&faint LE&-6&0&L\\
24$^{P,T,X}$&2013-04-11 07:24&2013-04-11 08:24&2013-04-11 10:30&84&90&H\\
25$^{NM,X}$&2013-06-23 22:36&2013-06-24 02:48&faint LE&59& &H\\
26$^{P,T,X}$&2013-07-09 15:12&2013-07-09 16:24& faint LE&12& &L\\
27$^{P,T,X}$&2014-08-15 17:48&2014-08-15 20:24&faint LE&-52& &H\\
28$^{X}$&2015-11-04 14:48&2015-11-04 15:24&2015-11-04 17:30&23&37&L\\
29$^{X}$&2016-10-09 02:24&2016-10-09 06:24&2016-10-09 10:18&-15&-35&L\\
30$^{X}$&2017-05-23 05:00&2017-05-23 08:24&2017-05-23 13:29&15&-3&L\\
31$^{X}$&2018-03-06 01:25&2018-03-06 03:48&faint LE&20& &L\\

\hline\hline

\hline                
\end{tabular}
\caption{Remote features of the observed CMEs. The first column is the event number with the indication of where the CME-ICME association was taken from and is followed by the CME's first C2 appearance time. The third column corresponds to the time the ellipse fit was performed in LASCO-C2 FOV, and the fourth column is the time the ellipse fit was performed in LASCO-C3 FOV. The fifth and sixth columns show the tilt results derived from LASCO-C2 and LASCO-C3, respectively. The last column shows the dominant inclination obtained from Tilt C2 and Tilt C3 values (see text for details); "L" stands for low inclination, "H" stands for high inclination, and "LE" stands for the leading edge.} 
\end{table*}
%

In Table \ref{table:1}, one can see that the inclination derived from LASCO-C2 may differ from the inclination derived from the LASCO-C3 coronagraphic images. The CME evolves through the entire FOV of C2 and C3, and by marking slightly different leading edges (green crosses in Figure \ref{Fig3}) at different times, we can infer slightly different inclinations for the same event. We note that this is not necessarily related to strong rotations and deflections in the LASCO-C2 or LASCO-C3 FOV (\citealp{Yurchyshyn2009}; \citealp{Vourlidas2011}; \citealp{Kay2017}) but to simple ambiguities inherent in the measurements. This is also visible in Figure \ref{Fig3}, where in LASCO-C3 FOV the ellipse is slightly less inclined than in the LASCO-C2 FOV. This is one of the reasons why we focus only on the dominant inclination. 

\subsection{Sheath region nonradial flows and shock orientation}

\noindent The boundaries of the MO and sheath region were determined manually for each event. We note that the selection of ICME boundaries involves a degree of uncertainty. In the first instance, the boundaries of the MO were chosen to cover the entire magnetic field rotation. When this was not possible due to the rotation of several magnetic field components, the events were excluded. As mentioned earlier, there were 30 events where this was the case. From left to right, the columns in Table \ref{table:2} show the event number, the date of the MO onset, shock-clear sheath occurrence time $SH_{\rm start}$, clear sheath end time $SH_{\rm end}$, the MO onset time, the MO end time, the derived FR type, the NRF ratio, the shock orientation $\theta_B$, the observed transit time TT, and $\gamma$ parameter. The sheath region was divided into two parts in some cases. The first part is the region where only clear sheath signatures can be seen (i.e., a strongly fluctuating magnetic field and plasma with increased density, temperature, and plasma beta). The second part of the envelope has fewer high plasma parameters and/or a not as strongly fluctuating magnetic field. This part shows no clear sheath and no clear MO properties. We identified this second part in 14 out of 31 events, as shown in Table \ref{table:2} (see column $SH_{\rm end}$).  In these 14 events, the end of the clear sheath region does not correspond to the beginning of the MO part. This part between the clear sheath and the clear MO was studied by \cite{Kilpua2013}, who recognized it as the disturbed front part of the FR known as the MO front region. More recently, \cite{Temmer2022} recognized this as compressed ambient solar wind and noted it as a leading edge structure. An example of a sheath with clear sheath properties is shown in the left panels of Figure \ref{Fig2}, while an example of a more complex sheath where the clear sheath is observed after the shock but then toward the MO part of the ICME one can also see a region with both sheath and MO properties is shown in the right panels of Figure \ref{Fig2}. There, one can observe a region that shows a stronger magnetic field with fewer fluctuations than in the clear sheath part. The density and plasma beta parameter show a further increase accompanied by a decrease in the temperature.    \\

%
\begin{table*}           
\label{table:2}      
\centering          
\begin{tabular}{ l l l l r r r l l r r}     
 Nr. & In Situ Date & $SH_{\rm start}$ & $SH_{\rm end}$ & $MO_{\rm start}$  & $MO_{\rm end}$ & FR type & NRF ratio & $\theta_{B}[^\circ]$ & TT[h] & $\gamma$[10$^{-7}$ km$^{-1}$]\\
\hline\hline
1&1997-01-10&10.04& &10.21&11.14&SWN&0.56&51&46.46&0.096\\
2&1997-10-10&283.68&283.92&284.15&285&SWN&0.88&89&98.33&8.901\\
3&1997-11-07&310.95&311.26&311.68&312.57&WNE&1.85& no data&78.4&0.431\\
4&1998-01-07&6.58&6.98&7.11&8.4&ENW&1.28&59&90.31&0.418\\
5&2000-08-12&224.8& &225.25&226.25&SEN&1.02& 64&59.92&0.125\\
6&2000-11-06&311.4&311.55&311.95&312.65&SEN&1.06& 46&68.3&1.141\\
7&2001-04-28&118.2&118.48&119.08&119.6&SEN&1&48&58.34&0.460\\
8&2002-03-19&77.55& &78.24&79.52&WNE&0.96&39&75.59&0.355\\
9&2002-04-17&107.47&107.7&108.02&109.15&SWN&0.92&66&64.23&0.137\\
10&2003-08-18&229.58& &230.12&231.25&ESW&1.13&62&53.9&2.332\\
11&2005-05-15&135.12&135.26&135.4&136.1&ENW&2.39&62&38.58&0.180\\
12&2008-12-17&351.5& &352.2&352.8&NWS&1.22& no data&102.34&4.782\\
13&2010-04-05&95.35&95.48&95.53&96.57&NWS&0.43&54&45.31&\\
14&2010-06-21&171.95& &172.35&173.7&NES&1.76& no data&99.07&4.169\\
15&2011-06-05&155.85& &156.05&156.42&WNE&0.98&69&61.7&0.239\\
16&2011-09-17&260.15&260.35&260.69&261.49&SEN&1.02&87&80.39&0.227\\
17&2011-10-25&297.78&297.91&298.05&298.67&ENW&0.57&64&66.03&0.106\\
18&2012-01-22&22.25& &22.52&22.77&NWS&0.87&85&66.73&0.465\\
19&2012-06-16&168.84&168.95&169.05&169.51&NES&1.76&60&55.95&0.135\\
20&2012-07-15&196.77& &197.3&199.1&ESW&1.09&33&70.28&0.627\\
21&2012-10-08&282.22& &282.76&283.35&ESW&1.84&74&81.32&0.502\\
22&2012-11-13&317.97& &318.4&319.15&NES&0.66&68&84.6&0.276\\
23&2013-01-17&17& &17.71&18.5&SWN&0.58& no data&88.54&7.762\\
24&2013-04-14&103.95& &104.75&105.95&ENW&2.5&40&78.91&1.118\\
25&2013-06-28&178.6& &179.1&180.5&WSE&2.23&74&88&0.831\\
26&2013-07-13&193.72& &194.25&195.35&NWS&1.14&79&78.43&0.087\\
27&2014-08-19&231.28&231.77&231.9&233.48&WNE&1.2&85&90.06&\\
28&2015-11-07&310.75&311.08&311.3&312.48&SWN&0.82&46&59.28&0.222\\
29&2016-10-13&286.92& &287.25&288.62&SEN&1.21&16&68.47&0.195\\
30&2017-05-28&147.65&147.9&147.98&149&SWN&1.33&81&101.1&0.065\\
31&2018-03-10&68.75& &69&69.8&SWN&0.42& no data&54.89&0.164\\

\hline\hline

\hline                
\end{tabular}
\caption{ In-situ derived features of ICMEs, shock angle $\theta$, and $\gamma$ parameter obtained with the reverse modelling procedure. First column shows the event number. Next is the date of MO onset followed by sheath onset time ($SH_{\rm start}$); sheath end time ($SH_{\rm end}$); MO onset time ($MO_{\rm start}$); and MO end time ($MO_{\rm end}$), all given in day of the year (DOY). The following columns show the FR type, NRF ratio, shock orientation $\theta_B$, observed transit time (TT) in hours. Finally, the gamma parameter is given in the last column.}
\end{table*}
%
Interplanetary CMEs are usually associated with NRFs in (1) the sheath region and (2) the expanding magnetic ejecta part. The first association is due to the plasma motion of the ambient solar wind escaping around the ICME ejecta part, and the second is related to the expansion of the magnetic ejecta in the nonradial direction, as described in \cite{Al-Haddad2022}. The NRF in the sheath region was previously studied by \cite{Gosling1987}. They discovered a westward flow related to the magnetic stress of the Parker spiral acting on ICMEs. Later, \cite{Owens2004} showed that the NRF in the sheath region can be used as an indicator of the local axis orientation of ICMEs and the point at which spacecraft and ICMEs meet. Additionally, \citet{Liu2008} investigated whether NRFs in the sheath could relate to the curvature of the MO. 

Similarly, \cite{Martinic2022} showed how differently oriented ICMEs may have different NRFs. We calculated the NRF ratio between the plasma flow in the $y$ and $z$ directions of the GSE coordinate system. The NRF flow is defined as the average of the absolute flow of the plasma in the $y$ or $z$ direction in GSE. The NRF ratio for each event is given in Table \ref{table:2}, column 8. We emphasize that the NRF ratio was determined from the part of the sheath where we observed only unique sheath features. For the 14 events mentioned above with complex sheath structures, this means that only the first part of the sheath was considered. In addition to the NRF in the sheath region, the shock orientation $\theta_B$, that is, the angle between the shock normal vector $\hat{n}$ and the upstream magnetic field $B_{up}$:

\begin{equation}
    \theta_B=\frac{180^\circ}{\pi}\arccos\Big(\frac{|B_{up}\cdot\hat{n}|}{||B_{up}||\:||\hat{n}||}\Big).
\end{equation}

The shock normal vector $\hat{n}$ was calculated by the mixed-mode method \cite{Abraham-Shrauner1976}, and in the cases where the data gap of velocity components was present, magnetic coplanarity from \cite{Colburn1966} was used. (For more detail on the $\hat{n}$ calculation, we refer the reader to the database of interplanetary shocks from which the $\theta_B$ were obtained.\footnote{\url{http://ipshocks.fi/database}}). The shock orientation $\theta_B$ values are given in Table \ref{table:2}. One can notice that not all events from Table \ref{table:2} have a corresponding $\theta_B$. These events (3, 12, 14, 23, and 31) do not meet the shock criterion given in the database of interplanetary shock documentation. However, they have a sheath developed enough to compute NRFs, as indicated above.

\subsection{Transit time}

 The transit time (TT) was calculated as the time difference between the time of onset of the ICME MO in the in situ data and the CME start time at 20~R$_s$ (solar radii). We note that this transit time is not the same as the one typically given in databases that corresponds to the arrival time of the shock. The CME start time at a starting radial distance of 20~R$_s$ was taken from the second order fit of the altitude-time measurements provided by SOHO/LASCO CME catalog.\footnote{\url{https://cdaw.gsfc.nasa.gov/CME_list/}} When measurements were only available for starting radial distances less than 20~R$_s$, an interpolation was performed using the acceleration corresponding to the same second order fit. 

\subsection{Drag-based model and $\gamma$ parameter determination}
\label{sec2.4}

Observational studies have derived that drag force dominates ICME propagation after a certain distance in the heliosphere. Results from these studies have formed the basis of numerous drag-based CME models (\citealp{Vrsnak2013}; \citealp{Hess2015}; \citealp{Mostl2015}; \citealp{Kay2018}), which apply the simple analytical equation:

\begin{equation}
    F_d=\gamma (v-w)|v-w|,
\end{equation}

\noindent where $v$ is the CME velocity, $w$ is the solar wind velocity, and $\gamma$ is the so-called drag parameter given by the following equation \citep{Vrsnak2013}:

\begin{equation}
    \gamma=C_d\frac{A\rho_w}{M+M_V}.
\end{equation}

\noindent Here, A is the cross-sectional area of the CME, $\rho_w$ is the solar wind density, $M$ is the CME mass, $M_V$ is the mass corresponding to the volume of the fluid displaced by the movement of the body (the so-called virtual mass), and $C_d$ is the dimensionless drag coefficient. We emphasize that $C_d$ is usually taken as one and as a constant during the propagation of the ICME. However, \cite{Cargill2004} has shown that the value of $C_d$ depends on the relative density and velocity of the CME with respect to the density and velocity of the solar wind. Cargill also showed that the value of $C_d$ increases from one for dense CMEs to as high as three for low-density CMEs and that $C_d$ has a significant radial dependence for the latter.

The drag parameter $\gamma$ is a very important parameter in the context of the drag force acting on a CME. Due to its dependence on CME cross section, mass, virtual mass, and solar wind density, obtaining the drag parameter $\gamma$ through direct measurements is currently unreliable \citep[see e.g.][]{Vrsnak2013,Dumbovic2021}. To derive the most reliable gamma value for our data sample, we used a reverse modeling method with the drag-based ensemble version v3 tool  \citep[DBEMv3 tool;][] {Calogovic2021}. In DBEMv3, input parameters (CME start time, CME source region longitude, CME half-width, solar wind speed, starting speed of CME, and $\gamma$ parameter)  with their uncertainties follow a normal distribution, with the observation input value set as the mean and three standard deviations as the uncertainty. The DBEMv3 tool creates 100,000 ensemble members from these input parameters and performs a single DBM run for each of them. For more detail on the creation of ensemble members using the DBEMv3 tool, the reader is referred to \cite{Calogovic2021}, and for a comprehensive description of the basic DBM and later developed versions, such as this ensemble version, to \cite{Dumbovic2021}. The reverse modeling method with DBEM has also been used by \cite{Paouris2021} to find the optimal $\gamma$ parameters and solar wind speed for a different subset of CME-ICME pairs.

For this particular study, the input parameters of CME start time, CME source region longitude, and CME half-width were set without uncertainties. These values are given in Table \ref{table:3}. The derivation of the CME start time is described in Sect. 2.3. The CME source region was determined from low coronal signatures: post-flare loops, coronal dimmings, sigmoids, flare ribbons, and filament eruptions.  For this, we used the JHeliowiever \citep{Muller2017} visualization tool. We analyzed 171, 211, 193, and 304 $\AA$ filtergrams from SDO/AIA \citep{Lemen2012} and SDO/HMI \citep{Scherrer2012} magnetogram data. When these data were not available, we used SOHO/EIT \citep{Delaboudini1995} and SOHO/MDI \citep{Scherrer1995} magnetogram data. The CME half-width, $\lambda$, was set to $89^\circ$ because all events were (partial) halo events as seen in the LASCO-C2 and LASCO-C3 FOV. The solar wind speed $w$ and the starting speed of CME $v_0$ follow a normal distribution, with the mean value being an observed value given in Table \ref{table:3}. The solar wind speed was obtained from in situ plasma measurements provided by the OMNI database \cite{King2005}, and it was determined as the mean velocity of the solar wind over an undisturbed period of several hours prior to the arrival of the CME shock. The CME start speed was taken as a second order speed given in SOHO/LASCO CME catalog.\footnote{\url{https://cdaw.gsfc.nasa.gov/CME_list/}}  The uncertainty (i.e., 3$\sigma$ value) for both the CME start speed and solar wind speed was set to 10\% of the mean value. For the purpose of reverse modeling with DBEMv3, we set the allowed gamma range to  0.01-10 $10^{-7}$ km$^{-1}$ with an equal probability for all $\gamma$ parameters in this range (i.e., the $\gamma$ parameter followed a uniform distribution in this range). As part of the reverse modeling procedure, we searched for the optimal $\gamma$ parameters where the forecast transit time is within one hour of the actual observed transit time. The median values of these obtained $\gamma$ parameters are listed in Table \ref{table:2}.

Events 1, 10, 26, 27, 29, and 31 in Table \ref{table:3} are marked with an asterisk. For these events, the original DBEMv3 input was changed because there were no transit times matching the observed transit time within one hour (i.e., no $\gamma$ parameters were found). We studied those events in more detail, and we found that for events 1, 10, 29, and 31, the radial takeoff distance needed to be changed. For events 26 and 27, the takeoff speed and speed uncertainty needed to be increased.

The height at which the drag force begins to dominate is not universal and varies greatly from event to event (\citealp{Vrsnak2001}; \citealp{Sachdeva2015}; \citealp{Sachdeva2017}). For events 1, 10, 29, and 31, we found that a starting radial distance of $20 R_{s}$ is not suitable as a DBEM input because the CME is still accelerating at this distance, and its propagation is therefore not dominated by the drag force. To improve our input for these events, the starting distance was increased by the trial-and-error method until a suitable initial distance was found that provided a "perfect transit time" (similar to \citealp{Sachdeva2015}). For events 1, 10, and 31, this distance was found to be 70 $R_s$, and we found it to be 50 $R_s$ for event 29.

For events 26 and 27, we found that the initial CME speed at 20 $R_{s}$ may be underestimated. This speed underestimation might come from the use of the second order fit of the height-time measurements. The second order fit shows a very small deceleration in the LASCO FOV. A linear fit yielded slightly different velocity estimates that provided physical solutions to find an optimal $\gamma$ with DBEM for event 26. The uncertainties of the CME launch speed were also increased to 20$\%$ in order to better compensate for the initial underestimation of velocity. For event 27, even after considering the linear speed and after increasing the uncertainties of the initial velocity, the optimal $\gamma$ parameter was not found. It could be that the DBM does not capture the physics of this event well. The same is true for event 13. This CME was launched on 3 April 2010 and is a well-studied event (\citealp{Rodari2018}; \citealp{Zhou2014}; \citealp{Rollett2012}; \citealp{Temmer2011}; \citealp{Liu2011}). \cite{Temmer2011} reported quite complex CME dynamics in the LASCO FOV and later in the heliosphere. This CME initially strongly accelerated up to 1100 km s$^{-1}$ and then had an abrupt deceleration down to 800 km s$^{-1}$ (all below 20 $R_{s}$). Later, the CME again accelerated and decelerated in the heliosphere, possibly due to a high-speed stream crossing. Due to its complex dynamics, this event is not suitable for reverse modeling with the DBEM or DBM in general. We find that it is also important to emphasize that even more sophisticated 3D MHD models such as ENLIL were not able to correctly represent the propagation of this CME \citep{Temmer2011}.

We note that some of the obtained $\gamma$ values lay outside of an expected range, 0.2-2 10$^{-7}$ km$^{-1}$, as given by \cite{Vrsnak2013}. This is most prominent for events 2, 12, 14, and 23 (see Table \ref{table:2}). We also emphasize that such high $\gamma$ values might be unreal, but testing such an assumption is beyond the scope of this paper. This would require meticulous analysis of the pre-eruption state of the heliosphere as well as detailed eruption analysis (see \citealp{Zic2015} and \citealp{Temmer2012}). We also highlight that from a theoretical point of view (see Equation 2), for cases when the CME launch speed is close to the solar wind speed, the corresponding optimal $\gamma$ obtained by the reverse modeling with drag-based models can easily take on very large values that may not be physically plausible. However, we also note that the reverse modeling procedure gave results close to the expected range of values for the majority of events, (i.e., for 25 out of 31 events).

\begin{table}
\centering
\begin{tabular}{l c c c c c}
\hline\hline
 Nr. & $t_0$ & $v_0$  [kms$^{-1}]$ & $\phi_{\rm CME}$ $[^\circ]$& $w$ [kms$^{-1}]$ \\
\hline
1*&1997-01-08 06:34&625&6&375\\
2&1997-10-07 01:16&620&0&407\\
3&1997-11-04 09:55&700&25&335\\
4&1998-01-03 08:20&515&40&309\\
5&2000-08-09 21:05&720&-15&416\\
6&2000-11-04 02:30&643&0&475\\
7&2001-04-26 15:35&1084&20&444\\
8&2002-03-16 02:10&917&7&293\\
9&2002-04-15 08:15&731&7&331\\
10*&2003-08-15 15:14&630&0&471\\
11&2005-05-13 19:00&1689&0&415\\
12&2008-12-12 20:47&432&-10&339\\
13&2010-04-03 15:10&661&20&509\\
14&2010-06-17 05:20&397&2&370\\
15&2011-06-02 11:30&996&6&337\\
16&2011-09-14 08:10&457&7&413\\
17&2011-10-22 07:10&663&25&323\\
18&2012-01-19 17:45&1390&-24&326\\
19&2012-06-14 17:15&983&-7&297\\
20&2012-07-12 18:55&2265&-20&326\\
21&2012-10-05 08:55&804&15&318\\
22&2012-11-09 21:00&603&-20&284\\
23&2013-01-14 00:33&339&-22&403\\
24&2013-04-11 11:05&819&-15&390\\
25&2013-06-24 10:10&513&40&373\\
26*&2013-07-09 23:34&450&-20&386\\
27*&2014-08-16 03:00&342&5&295\\
28&2015-11-04 19:55&708&5&465\\
29*&2016-10-10 09:31&495&-17&355\\
30&2017-05-23 16:30&367&0&303\\
31*&2018-03-07 17:06&538&-8&366\\
\hline
\end{tabular}
\caption{DBEM input parameters. The number of the CME is indicated under Nr.; $t_0$ is the CME start date and time in UT at 20 $R_{\rm SUN}$; $v_0$ is the CME start speed at 20 $R_{\rm SUN}$ given in km\,s$^{-1}$; $\phi_{\rm CME}$ is the longitude of the CME source position in degrees; and $w$ is the solar wind speed in km\,s$^{-1}$.}
\label{table:3}
\end{table}

\section{Results and discussion}
\label{Results and discussion}

   \begin{figure*}
   \centering
   \includegraphics[width=2\columnwidth]{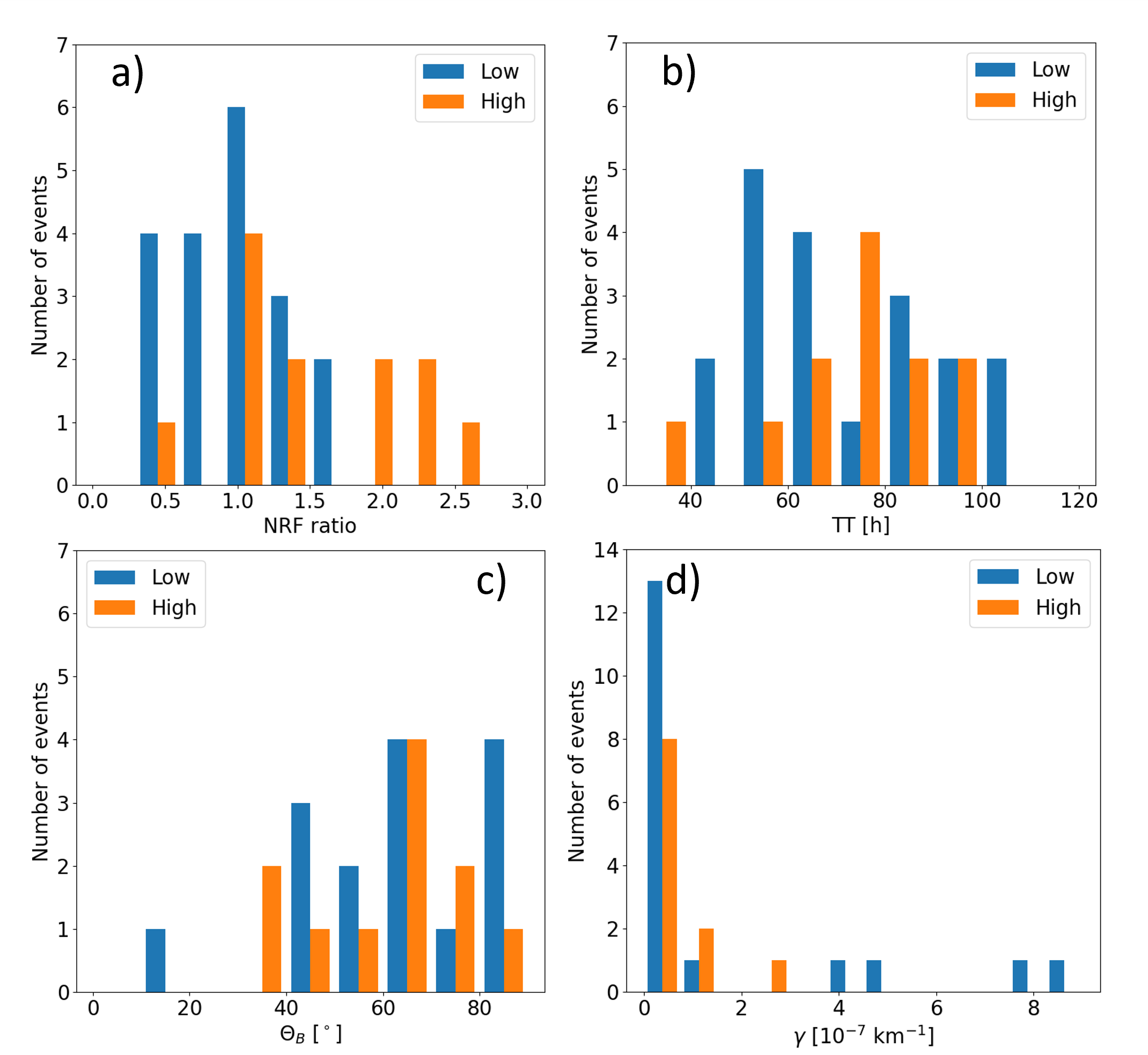}
   \caption{Distributions for NRF ratio, transit time (TT), shock orientation ($\theta_B$), and drag parameter $\gamma$ for high-inclination events (orange) and low-inclination events (blue).}
              \label{Fig4}
    \end{figure*}

%
\begin{table*}
\caption{Statistical results. Mmean, median, standard deviation, and 5. and 95. percentiles for low- and high-inclination events (reported separately).}             
\label{table:4}      
\centering          
\begin{tabular}{|r|c c c c|c c c c|}
\hline
      &       & \multicolumn{2}{c}{LOW INCLINATION} & \multicolumn{5}{c|}{HIGH INCLINATION} \\
      \hline\hline
     & NRF ratio & $\theta_B[^\circ]$ & TT[h] & $\gamma$[$\times 10^{-7} km^{-1}$] & NRF ratio & $\theta_B[^\circ]$ & TT[h] & $\gamma$[$\times 10^{-7} km^{-1}$]\\
 MEAN & 0.98 & 62.67 & 72.7 & 1.63& 1.5 & 60.09 & 72.8 & 0.65\\ 
 MEDIAN & 1.00 & 64 & 68.3 & 0.22 & 1.24 & 62 & 76.99 & 0.43\\
 STD &  0.37 & 19.31 & 18.63 & 2.80 & 0.61 & 15.64 & 15.02 & 0.60\\
 PERC[5,95] & [0.42,1.76] & [37,87.6] & [46.35,101.22] & [0.08,7.93] & [0.78,2.44] & [36,79.5] & [47.00,90.44] & [0.14,1.72]\\
 \hline 
\end{tabular}
\end{table*}
%

Dominant inclination results obtained from remote and in situ data are given in the last column of Table \ref{table:1} and the sixth column of Table \ref{table:2}, respectively. In Figure \ref{Fig4}, we show the occurrence frequency of dominantly low- and high-inclined events with respect to NRF, transit time, shock orientation, and $\gamma$ parameter. One can see that most of the high- and low-inclination events have NRF ratios close to one. However, there is a greater number of low-inclination events with low NRF ratios and a greater number of high-inclination events with high NRF ratios. This is consistent with the results of \cite{Martinic2022}, where a similar procedure was applied to a smaller sample of events. This suggests that NRFs are more pronounced in the $\pm y$ direction for events with high inclination and in the $\pm z$ direction for events with low inclination. The mean, median, standard deviation, and 95th percentile for NRF ratios are shown in Table \ref{table:3}. The mean, median, and 95th percentile show larger values for high-inclination events, confirming the results of the distribution plot in panel (a) of Figure \ref{Fig4}. We observed that the standard deviation for high-inclination events is almost twice the standard deviation of low-inclination events, which is related to the spread of NRF values. Namely, low-inclination events can be found in the 95th percentile interval $[ 0.42, 1.76 ]$, while high-inclination events have a 95th percentile interval $[ 0.78, 2.44 ]$. 

As stated earlier, the NRF ratios were calculated from the velocity in the $y$ and $z$ directions of the GSE coordinate system in the clear sheath part of the ICME and are a consequence of ambient plasma interacting with the FR part of the ICME. However, we note that the deflection of plasma due to fast-forward shock may also contribute to the NRF and this contribution cannot be easily disentangled from the contribution due to draping. In order to confirm that the above-stated dependence of NRF ratios on ICME inclination comes from plasma being deflected around the ICME FR part rather than from plasma that is being deflected on the shock front, we calculated the shock orientation and studied the dependence of shock orientation on inclination. This dependence can be seen in the distribution of $\theta_B$ in panel (c) of Figure \ref{Fig4}. Unlike NRF ratios, the shock orientation (which determines the shocked plasma deflection right behind the shock front) does not show dependence on ICME inclination. From Table \ref{table:2}, we also observed that most events have $\theta_B$ greater than 45$^\circ$, which means that most of the events studied have a quasi-perpendicular shock front.

 In order to quantitatively test the difference between low- and high-inclination samples, we performed the Welch's test (in case of different sample variances) and the student t-test (in case of similar sample variances). First, in order to choose an adequate test for the means of the populations, we had to test the sample variances. To see whether two samples have similar or different variances we used a statistical F-test. According to the F-test, with a 95$\%$ confidence level, the shock orientation $\theta_B$ and transit time have similar variances for high- and low-inclination groups of events; however, for NRF ratio and gamma parameter $\gamma$, these two groups of events show statistically significant variances. High-inclination events (orange bars in Figure \ref{Fig4}) have a wider spread in NRF ratios in comparison to low-inclination events (blue bars), shifting the distribution toward higher NRF ratio values. Regarding the $\gamma$ parameter, low-inclination events (blue bars in Figure \ref{Fig4}, panel d) have a wider spread. The same is not valid for transit time and shock orientation.

 Welch's test null hypothesis is that the NRF ratios for low- and high-inclination events come from random samples from normal distributions with equal means and unequal variances. Welch's test was performed under the assumption that
(1) the NRF ratio/ $\gamma$ parameter for high- and low-inclination events are independent, (2)\ the NRF ratio/ $\gamma$ parameter distributions for low- and high-inclination samples are normal, and (3) the NRF ratio/ $\gamma$ parameter variances for low-inclination and high-inclination events are different (according to the F-test).

The result of Welch's test for NRF ratios is that the null hypothesis should be rejected at the 95$\%$ significance level (i.e., the NRF ratios for high- and low-inclination events come from populations with unequal means). The interpretation of the different NRFs observed for ICMEs with different inclinations comes from the fact that the ambient plasma in front of the ICME bypasses the obstacle (ICME FR) in a way where the extent of the obstacle is smaller. For ICMEs with low inclination, the extent of the ICME FR part in the $\pm z$ direction is smaller than in the $\pm y$ direction, and therefore the NRF ratio is smaller for ICMEs with low inclination. In contrast, the extent of the ICME with high inclination is smaller in the $\pm y$ direction, so the plasma flows mainly in this direction. A sketch of the various NRFs in terms of the different inclinations of CMEs  is shown in \cite{Martinic2022}. The result of Welch's test for the $\gamma$ parameter is that the null hypothesis should not be rejected (i.e.,  the $\gamma$ parameter for high- and low-inclination events comes from populations with equal means). Welch's test is based on the normality assumption, which is hardly satisfied for $\gamma$ values (see histogram in Figure \ref{Fig4}, panel d). The Kolmogorov-Smirnov test and Mann-Whitney U-test, as nonparametric significance tests, were also performed. However, we note that both tests confirmed the results from Welch's test at the same confidence interval (95\%), meaning that there is no significant difference between low- and high-inclination events regarding $\gamma$ values.

For shock orientation and transit time, the F-test confirmed similar variances for low- and high-inclination samples. Thus, instead of Welch's test, the student t-test was performed under the assumption that
(1) the shock orientation/transit time for high- and low-inclination events are independent, (2) the shock orientation/transit time distributions for low- and high-inclination samples are normal, and (3) the shock orientation/transit time variances for low-inclination and high-inclination events are similar (according to the F-test).

The t-test confirmed the null hypothesis at the 95$\%$ significance level, meaning that the samples of shock inclination and transit time for low- and high-inclination events come from populations with equal means. In other words, there is no statistically significant difference between low- and high-inclination groups of events.

 The fact that there is no difference in the $\gamma$ parameter and transit time for differently oriented CMEs suggests that the orientation of the CME does not affect the overall drag of the CME. However, we note that the drag depends primarily on the difference between the velocity of the CME and the ambient solar wind speed. In addition, the $\gamma$ parameter depends on the CME cross section, the ambient solar wind density, the mass of the CME, and the virtual mass. It is possible that the effect of inclination is small enough to be "masked" by all these contributions, even though we selected the sample in order to minimize them. 
As described in \cite{Martinic2022}, the inclination effect on the drag should be most pronounced at the minimum of the solar cycle, where the configuration of the IMF most closely matches that of a simple magnetic dipole. While our sample of events includes some that occurred near the minimum of solar activity (event numbers 11,12,13,14, and 31), the majority of events correspond to the maximum, when the IMF configuration is very complex. Due to the very small sample of events at the minimum of solar activity, no analysis of the difference between events at the minimum and maximum of activity was performed.

Except for inclination influence, \cite{Vandas1995} and \cite{Vandas1996} also emphasized the importance of the chirality of the CME for its propagation, which is not captured by our study. This was later tackled by \cite{Chane2006}, who studied the propagation of two CMEs: one in which the initial magnetic field and the background magnetic field had the same polarity and another where they had opposite polarities. Their simulations showed that the initial magnetic polarity significantly affects the evolution of CMEs. We note here that the study of \cite{Chane2006} did not examine the effects of CME inclination but rather the effects of initial chirality on propagation in the inner heliosphere. More recently, \cite{Shen2021} studied the effects of different initial CME densities, masses, sizes, and magnetic field configurations on simulation results for observers near Earth and Mars. Nevertheless, to our knowledge, there are no 3D MHD studies aimed specifically at investigating the effects of (I)CME inclination and its interaction with the environment, such as IMF draping and plasma flows ahead of the ICME. Such a study could beneficially complement our findings based on observations.

\section{Summary and conclusions}
\label{Summary and conclusions}

Altogether, 31 Earth-directed CME-ICME pairs with distinct magnetic obstacle (MO) properties and pronounced sheath regions during the period from 1997 to 2018 were studied. We inferred the dominant inclination from the ellipse fitting of LASCO-C2 and LASCO-C3 coronagraphic images. The dominant inclination was also derived from in situ data of the rotation of magnetic field components in the MO part of the ICME. Of the 31 CME-ICME pairs, 19 are low-inclination events, and 12 are high-inclination events.

Some basic features of the ICME propagation in terms of the inclination of the event were analyzed. We investigated the NRFs in the sheath region along with the shock orientation, transit time, and $\gamma$ parameter. We found a significant difference in NRFs for differently oriented ICMEs. Low-inclination events were found to have lower NFR ratios, while high-inclination events were found to have higher NFR ratios. This implies that low-inclination events are more likely to have ambient plasma escape via the meridional plane, while high-inclination events are more likely to have plasma escape via the ecliptic plane \citep[see ][]{Martinic2022}.

The plasma deflection on the fast-forward shock could also contribute to the measured NRF ratios. To confirm that the above-stated difference between low- and high-inclination events is indeed due to the deflection of the plasma around the obstacle (ICME FR part) and not due to the deflection of the plasma by the shock front, we examined the dependence of the NRF ratios on the shock orientation. We found no differences in the NRF occurrence frequency with respect to the shock orientation, thus confirming the result stated above.

No significant difference was found in the transit time and $\gamma$ parameter for differently oriented ICMEs. This suggests that the predominant inclination of the ICME has no effect on the drag due to the interaction with the ambient solar wind and IMF. We note that by inclination we mean tilt, that is, the angle between the elliptic plane and ICME flux rope axis, not the magnetic field orientation. We also emphasize that most of the studied events occurred near solar maximum, which is when the IMF has a very complex configuration. It is also possible that the influence of the inclination on the drag force is much smaller than the contributions of other features, such as the difference between the speed of the CME and the solar wind, the CME mass, the CME cross section, and the ambient density, and therefore the inclination effect is very difficult to decipher.

\begin{acknowledgements}
      We acknowledge the support by the Croatian Science Foundation under the project IP-2020-02-9893 (ICOHOSS).
      K.M. acknowledges support by the Croatian Science Foundation in the scope of Young Researches Career Development Project Training New Doctoral Students.
      N.~A. acknowledges grants NSF AGS1954983 and NASA-ECIP 80NSSC21K0463.
      We also acknowledge the support from the Austrian-Croatian Bilateral Scientific Projects ”Comparison of ALMA observations with MHD-simulations of coronal waves interacting with coronal holes” and ”Multi-Wavelength Analysis of Solar Rotation Profile”
      This paper uses data from the Heliospheric Shock Database, generated and maintained at the University of Helsinki
      The SOHO/LASCO data used here are produced by a consortium of the Naval Research Laboratory (USA), Max-Planck-Institut fuer Aeronomie (Germany), Laboratoire d’Astronomie (France), and the University of Birmingham (UK). SOHO is a project of international cooperation between ESA and NASA.
      We acknowledge use of NASA/GSFC’s Space Physics Dana Facility’s OMNIWeb (or CDAWeb or ftp) service, and OMNI data.
      
\end{acknowledgements}

%
%

%
%
%
%
%
%
%
%
%

   \bibliographystyle{aa} 
   \bibliography{REF_KM} 

\end{document}